1# On the Achievable Secrecy Diversity of Cooperative Networks with Untrusted Relays

Mohaned Chraiti, Ali Ghrayeb, Chadi Assi and Mazen O. Hasna*Abstract*—Cooperative relaying is often deployed to enhance the communication reliability (i.e., diversity order) and consequently the end-to-end achievable rate. However, this raises several security concerns when the relays are untrusted since they may have access to the relayed message. In this paper, we study the achievable secrecy diversity order of cooperative networks with untrusted relays. In particular, we consider a network with an $N$-antenna transmitter (Alice), $K$ single-antenna relays, and a single-antenna destination (Bob). We consider the general scenario where there is no relation between $N$ and $K$, and therefore $K$ can be larger than $N$. Alice and Bob are assumed to be far away from each other, and all communication is done through the relays, i.e., there is no direct link. Providing secure communication while enhancing the diversity order has been shown to be very challenging. In fact, it has been shown in the literature that the maximum achievable secrecy diversity order for the adopted system model is one (while using artificial noise jamming). In this paper, we adopt a nonlinear interference alignment scheme that we have proposed recently to transmit the signals from Alice to Bob. We analyze the proposed scheme in terms of the achievable secrecy rate and secrecy diversity order. Assuming Gaussian inputs, we derive an explicit expression for the achievable secrecy rate and show analytically that a secrecy diversity order of up to $\min(N, K) - 1$ can be achieved using the proposed technique. We provide several numerical examples to validate the obtained analytical results and demonstrate the superiority of the proposed technique to its counterparts that exist in the literature.

*Index Terms*—Cooperative networks, interference alignment, interference dissolution, secrecy diversity, untrusted relays.## I. INTRODUCTION

### A. Background

Owing to its great potential in improving the communication reliability of wireless networks, cooperative relaying has been adopted as a key technology in major wireless standards, including IEEE 802.11, among others [1]. One of the main advantages of cooperative relaying is that it increases the transmission range [2], which is crucial for achieving ubiquitous coverage without compromising the quality of service provided to customers [3]. Cooperative relaying will certainly continue to play a central role in shaping the backbone of next generation wireless networks. This can be clearly seen from the proliferation of heterogeneous devices connected through large-scale networks, including Ad-hoc network femto-cells, and the Internet of things (IoT), where information is delivered to remote destinations through relaying. While cooperative relaying has obvious benefits, it also poses serious security risks. Indeed, relays are usually decentralized and it is often the case that relays can be as simple as hand-held devices, with limited computational capability. Consequently, relying on cryptography-based techniques will be severely limited and may not be an option [4]. Even when it is possible to use cryptography, it may not be possible to achieve secure wireless communication. This is due to the fact that information in packet overheads is left unencrypted to facilitate communication, and such overheads could provide a wealth of information for eavesdroppers. This challenge has recently led to an overwhelming interest in physical layer security (PLS), which is considered as a first-line of defense against eavesdropping [5]–[7].

The concept of PLS was coined by Wyner in his seminal work [8] in which he introduced the wiretap channel. He proved that, for a point-to-point degraded wiretap channel, where an eavesdropper (Eve) receives a degraded version of the signal sent from a transmitter (Alice) to a legitimate destination (Bob), secure communication can be provided without sharing a secret key between the communicating parties. This result is encouraging in the sense that it is possible to establish secure communications independent of how powerful eavesdroppers can be in terms of computational capability. However, the situation is completely different for cooperative networks, especially when there is no direct link between the transmitter and the legitimate destination. In such a scenario, the signal received at the legitimate destination is a degraded version of the signal received by the relays. It could be the case that it is desired to not allow relays receiving the signal to be able to extract any information contained by the signal. To achieve this, the relays involved in relaying should be considered as untrusted relays. Furthermore, one should assume that there could be eavesdroppers adjacent to the relays who will receive a copy of the signal received by the untrusted relays. To the end, it is easy to conclude that it is possible that the signal received at the destination is a degraded version of the one received by untrusted relays and eavesdroppers. This renders PLS-related techniques/results obtained for point-to-point wireless networks irrelevant. This necessitates coming up with new techniques to deal with this problem. Among the performance metrics that need to be revisited in light of the

M. Chraiti is with the ECE Department, Concordia University, Montreal, Canada (email:m_chrait@encs.concordia.ca).

A. Ghrayeb is with the ECE Department, Texas A&M University at Qatar, Doha, Qatar and with Qatar Computing Research Institute (QCRI), HBKU, Qatar (e-mail: ali.ghrayeb@qatar.tamu.edu).

C. Assi is with the CIISE Department, Concordia University, Montreal, Canada (email:assi@ciise.concordia.ca)

M. O. Hasna is with Electrical engineering Department, Qatar University, Doha, Qatar (e-mail: hasna@qu.edu.qa)

This work was supported by Qatar National Research Fund (a member of Qatar Foundation) under NPRP Grant NPRP8-052-2-029, by an NSERC Discovery Grant and Concordia University. The statements made herein are solely the responsibility of the authors.

above mentioned challenge include the achievable secrecy rate and secrecy diversity order, which are the focus of this paper.

*B. Relevant Literature*

The fundamental question concerning exploiting untrusted relays to enhance the secrecy rate was explored in [4], [9]–[13] with favorable results. The authors in [4] considered a cooperative system with an untrusted relay and they proposed artificial noise-based transmission schemes whereby the destination sends artificial noise (AN) while the transmitter sends the information signal (i.e., destination-aided jamming). Then the (untrusted) relay amplifies and forwards the combined signal (the information and AN signals). The destination uses its perfect knowledge of the AN to subtract it form the received signal, and then decodes the intended signal. This technique was extended in [9] to the case of multiple-input multiple-output (MIMO) systems with untrusted relays. The authors showed that it is beneficial in terms of the secrecy rate to treat relays as untrusted as opposed to treating them as eavesdroppers. However, the loss in the achievable secrecy rate, as compared to the case when the relays are trusted, could be significant.

It is somewhat intuitive to assume that the secrecy rate is improved as the number of untrusted relays increases. However, the authors in [10], [11] showed the opposite; that is, the secrecy rate decreases as the number of untrusted relays increases. Moreover, it was shown in [12] that the achievable spatial diversity associated with the confidential message is limited to one, regardless of the number of untrusted relays used in relaying. This is bad news because it suggests that the loss incurred by treating the relays as untrusted degrades the diversity significantly. As a remedy to this problem, the authors in [13] proposed a scheme that involves using inter-relay jamming whereby relays jam each other in an effort to keep information confidential. The system model considered in [13] comprises an $N$-antenna transmitter, $K$ single-antenna relays and a single-antenna destination, with the assumption that the number of transmit antennas is larger than the number of relays. It was shown that the achievable secrecy diversity is improved to $K-1$. However, while this is a positive result, it may not be useful because, in practical settings, the number of relays is normally much larger than the number of transmit antennas. Furthermore, the schemes in [10], [11] use some forms of AN.

*C. Problem Statement*

The techniques pertaining to the problem at hand that exist in the literature, as mentioned above, achieve at most a secrecy diversity order of one for the practical scenario when the number of relays is greater than the number of transmit antennas. This clearly limits exploiting the full potential of having many untrusted relays in cooperative networks. Moreover, the notion of using AN (whether relay-based or destination-based), which is essentially used in [4], [9], [10], [12], [13], may result in three main issues. First, in these works, AN is assumed to be treated as noise by the eavesdroppers and this stems from the belief that two interfering signals are indistinguishable in a one-dimensional space unless a signal is treated as noise while decoding the other. However, recent works showed that it is possible to jointly decode intended and interfering signals if they are transmitted over different channels [14]–[17]. For instance, in [14], we proposed a scheme that is able to break up a one dimensional space into two fractional dimensions. As such, a destination equipped with one antenna can perfectly extract two interfering signals received via different channels. In [15]–[17], the authors proposed techniques dealing with interference in a one-dimensional space, where they showed that it is possible to jointly decode intended and interfering signals received over different channels for almost all channel realizations, which proves that interfering signals are naturally aligned by the channel. They also showed that two signals belonging to a discrete constellation are inseparable only if they are transmitted simultaneously over the same channel, i.e., aligned by the same channel. Even when signals do not belong to discrete constellations, they can be discretized to make their effect less severe by applying real interference alignment. In light of these results, it is reasonable to assume that an eavesdropper, collocated with a relay, may have the possibility of efficiently decoding an intended signal in the presence of AN, rendering relying on AN-based techniques inefficient.

Second, AN-based approaches have the disadvantage of consuming power given that the relay amplifies and forwards the intended and AN signals at the same time. Third, when the number of untrusted relays is large, the destination, when acting as a jammer, will have to generate AN with high power to jam all the relays including the ones that are far from it. However, this results in the drawback of deteriorating the quality of the intended signal at the destination. To elaborate, considering that amplify-and-forward (AF) is used, the forwarded signal will comprise the intended signal and the AN signal. Given that the transmit power from the relays is limited, when the AN power is relatively large, the intended signal power will effectively be scaled down, which leads to a lower signal-to-noise ratio (SNR) at the destination (after the jamming signal is subtracted). Consequently, the achievable secrecy rate will deteriorate. This leads us to believe that relying solely on AN-based strategies may not be as efficient as one may think.

*D. Contributions and Outline*

Motivated by the above discussion, we propose in this paper a novel PLS scheme for cooperative networks with multiple untrusted relays. It is assumed that there is no direct link between the transmitter and the intended destination and all communication is done through the untrusted relays. This mimics a situation when the destination is far from the transmitter. The proposed scheme does not use any form of AN, which makes it completely different from its counterpart schemes that have been proposed in the literature. In developing the proposed scheme, we make use of an interference alignment scheme that we proposed in [14], which involves precoding signals in a nonlinear fashion with the channel gains such that only the destination will able to efficiently separate interfering signals.

The system model considered in this paper comprises an $N$-antenna transmitter, $K$ single-antenna untrusted relays and a single-antenna destination. We consider that $N, K \geqslant 2$. However, there is no additional constraints concerning the relation between $N$ and $K$, implying that $K$ may be arbitrarily large and hence greater than $N$. As often considered in the literature, we assume the worst case scenario where each relay is collocated with an eavesdropper that is able to get the same signal received by the relay it is collocated with. We analyze the performance of the proposed technique in terms of the achievable secrecy rate for which we drive a closed-form expression that explicitly shows the impact of the number of transmit antennas and number of untrusted relays on the achievable secrecy rate. We use the derived secrecy rate to derive the achievable secrecy diversity order. In particular, we invoke the notion of the secrecy outage probability to derive the achievable secrecy diversity order. Although the obtained expression is exact, it does not reveal explicitly the secrecy diversity order. To get around this, we derive an upper bound on the outage probability, which shows that the achievable secrecy diversity order is up to $(\min(N, K) - 1)$ while keeping the messages confidential with respect to all untrusted nodes. In contrast, the technique proposed in [12] offers secrecy diversity order one, and the one proposed in [13] is not applicable to the system model considered in this paper since $K$ can be larger than $N$.

Note that the proposed technique does not use AN, which helps to overcome the three issues related to the use of AN listed above. As a result, the proposed technique offers the advantage of significant power savings as compared to its counterpart schemes, and this directly translates by a higher transmission rate for the same total transmit power.

The rest of the paper is organized as follows. In Section II, we present the system model and we describe in details the proposed technique. In Section III, we analyze the performance of the proposed technique in terms of the achievable secrecy rate. In Section IV, we derive the achievable secrecy diversity order. Numerical results are provided in Section V. We conclude the paper in Section VI.

Throughout the rest of the paper, we use $|\cdot|$, $(\cdot)^*$, $(\cdot)^T$ and $\langle \cdot, \cdot \rangle$ to denote the 2-norm, the transpose conjugate, the transpose operators and the inner product between two vectors, respectively. We use $E[\cdot]$ to denote the expectation operator. In this paper, $I(\cdot)$ and $P_r[\cdot]$ denote the mutual information and the probability of an event respectively.

## II. PROPOSED SCHEME

### A. System Model

The system model considered in this paper is depicted in Fig. 1, which comprises an $N$-antenna transmitter (Alice), a single-antenna destination (Bob) and $K$ single-antenna relays denoted by $\{R_1, R_2, \ldots, R_K\}$. Knowledge of the whereabouts of Eves is unknown to the relays nor to Alice or Bob. Therefore, it is assumed that each relay is collocated with an eavesdropper (Eve). Each Eve receives the same signal received by the relay it is collocated with. This assumption is adopted to account for the worst case scenario in terms of the achievable secrecy rate, and it is commonly used in the literature. As it will be elaborated on later in Section III, among the available $K$ relays, $2 \leqslant L \leqslant \min(N, K)$ realys are selected at random to relay the confidential message to Bob. The remaining $K - L$ relays are treated as Eves. We are aware that there are other relay selection criteria available in the literature, but determining the optimal selection criterion with respect to some performance metric is out of the scope of this paper.

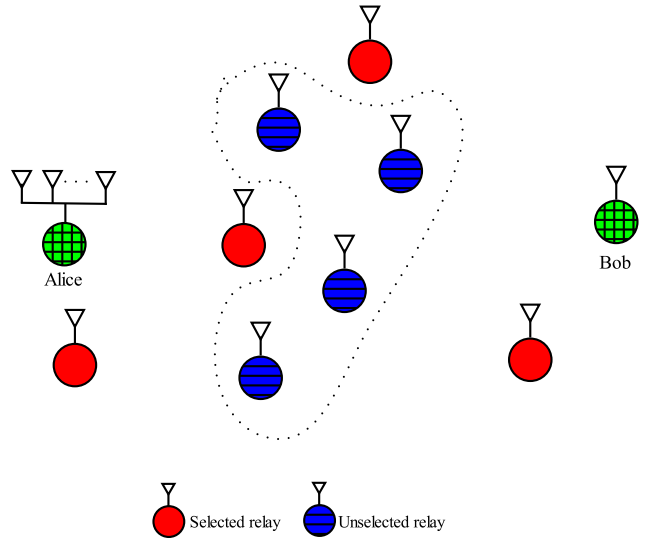

Fig. 1: System model.

All sub-channels are assumed to be quasi-static and the channel gains follow the Rayleigh distribution with variance one. The channel gains remain constant during a coherence time and change independently from one coherence time to another. For a given coherence time, the channel gains vector between the transmitter and $R_k$ is denoted by $\boldsymbol{h_k} = \{h_{k,1}, h_{k,2}, \ldots, h_{k,N}\}$. The channel gain between $R_k$ relay and Bob is denoted by $g_k$. We assume that all untrusted relays have all the channel state information (CSI) of all sub-channels which is the worst case scenario. Moreover, we assume that Alice has the CSI of the channel between Alice and the selected relays, and between the selected relays and Bob. The selected relays remain the same during a coherence time.

### B. Interference Dissolution for Cooperative Networks

We proposed in [14] an interference management technique, referred to as interference dissolution (ID), with the objective of managing interference in a one-dimensional space over time-invariant channels. We showed that ID can achieve a rate of two symbols per channel use. More importantly, it was shown that ID achieves a non-zero degree-of-freedom (dof) for all transmitted symbols, implying that all symbols are perfectly separable at the destination. Building on those promising results, we borrow in this paper the core idea of ID and adapt it to the underlying system model, i.e., cooperative networks with multiple untrusted relays.

Without loss of generality, let us assume that Bob intends to transmit to Alice $mL$ pairs of symbols, namely,

($\{x_{1,1}, x_{1,2}\}, \{x_{1,3}, x_{1,4}\}, \ldots, \{x_{L,2m-1}, x_{L,2m}\}$). Note that the restriction on grouping the transmitted symbols into pairs stems from the fact that ID was developed to achieve a rate of two symbols per channel use [14]. The transmit power per symbol is assumed to be $P$, i.e., $E[|x_{l,i}|^2] = P \ \forall \{l, i\} \in [1, L] \times [1, 2m]$.

Note that the $mL$ symbols will be transmitted through the selected $L$ relays. These symbols are divided into $L$ sets, where each set contains $2m$ symbols. As it will be shown below, $2(mL + 1)$ channel uses are required to transmit the $2mL$ symbols. During the first channel use, the transmitter beamforms each sum of $m$ pairs of symbols in the direction of a selected relay while nulling it out in the direction of the remaining relays. This is performed through zero forcing beamforming (ZFBF) [18]. That is, the sum of the $l$th symbol pairs $\sum_{i=1}^{2m} x_{l,i}$ ($l \in [1, L]$) is beamformed in the direction of R$_l$. This is done by multiplying it by the $l$th column vector $\boldsymbol{b}_l$ of the pseudo-inverse matrix

$$(\boldsymbol{b}_1^T, \boldsymbol{b}_2^T, \ldots, \boldsymbol{b}_L^T) = \boldsymbol{H}^*(\boldsymbol{H}\boldsymbol{H}^*)^{-1},$$

where $\boldsymbol{H} = (h_{l,i})_{\{l,i\} \in [1,L] \times [1,N]}$ is the channel matrix. The $l$th row of the channel matrix is denoted by $\boldsymbol{h}_l$. We note here that the multiplication with a beamforming vector may result in a power amplification [18]. To deal with this, the transmitter normalizes each beamforming vector with respect to its norm, and therefore the signal received by $textR_l$ is expressed as

$$z_{l,1} = h'_{l,l} \sum_{i=1}^{2m} x_{l,i} + n_{l,1}, \quad (1)$$

where $h'_{l,l} = \frac{1}{\|\boldsymbol{b}_l\|} \boldsymbol{h}_l \boldsymbol{b}_l^T$ and $n_{l,1}$ is AWGN with zero mean and variance $\sigma^2$. We use $\boldsymbol{h}' = \{h'_{l,1}, h'_{l,2}, \ldots, h'_{l,L}\}$ to denote the vector resulting from beamforming $\{\sum_{i=1}^{2m} x_{1,i}, \sum_{i=1}^{2m} x_{2,i}, \ldots, \sum_{i=1}^{2m} x_{L,i}\}$ to R$_l$ ($l \in [1, K]$). In (1), the elements $h'_{l,j} = \frac{1}{\|\boldsymbol{b}_j\|} \boldsymbol{h}_l \boldsymbol{b}_j^T$ ($j \neq l$) do not appear because they are equal to zero due to ZFBF.

In the second channel use, the relays simultaneously amplify and forward their respective received signals. The average power available at each relay is assumed to be $2mP$. Each relay has to normalize the received signal before transmission. The $l$th relay normalizes the received signals by factor $\alpha_l = \sqrt{|h'_{l,l}|^2 + \frac{\sigma^2}{2mP}}$. Consequently, the signal received by Bob can be written as

$$y_1 = \sum_{l=1}^{L} \left( \frac{h'_{l,l} g_l}{\alpha_l} \sum_{i=1}^{2m} x_{l,i} + \frac{g_l}{\alpha_l} n_{l,1} \right) + n_1, \quad (2)$$

where $n_1$ is AWGN with zero mean and variance $\sigma^2$. The received signal can be written also in the following form.

$$y_1 = \sum_{l=1}^{L} g'_l \sum_{i=1}^{2m} x_{l,i} + n'_1, \quad (3)$$

where $g'_l = \frac{h'_{l,l} g_l}{\alpha_l}$ and $n'_1 = \sum_{l=1}^{L} \frac{g_l}{\alpha_l} n_{l,1} + n_1$ is AWGN with zero mean and variance $\sigma^2 \left(1 + \sum_{l=1}^{L} \frac{|g_l|^2}{\alpha_l^2}\right)$.

In the third channel use, the transmitter precodes $\{x_{1,1}, x_{1,2}\}$, according to the ID technique [14], in order to allow the destination to properly separate them from the other symbols. To this end, the transmitter calculates a dissolution factor $\beta_1$ by solving [14]

$$g'_1 x_{1,1} + \beta_1 g'_1 x_{1,2} = \sum_{l=1}^{L} g'_l \sum_{i=1}^{2m} x_{l,i}, \quad (4)$$

which gives

$$\beta_1 = 1 + \frac{g'_1 \sum_{i=3}^{2m} x_{1,i} + \sum_{l=2}^{L} g'_l \sum_{i=1}^{2m} x_{l,i}}{g'_1 x_{1,2}}. \quad (5)$$

As detailed in [14], the transmitter sends a nonlinear combination of $s_1$, $s_2$ and $\beta_1$ and hence, the received signal has the following form.

$$y_2 = g'_1 x_{1,2} - \beta_1 g'_1 x_{1,1} + n'_2, \quad (6)$$

where $n'_2$ is AWGN. The noiseless part of $y_2$ is beamed in this case in the direction of all selected relays. To guarantee that the used power is $2mP$, the transmitted signal is normalized by $\varrho = \sqrt{\frac{1}{2mP} E[|g'_1 x_{1,2} - \beta_1 g'_1 x_{1,1}|^2]} = \sqrt{\sum_{1}^{L} |g'_l|^2}$. The received signal by R$_l$ can be then written as

$$z_{l,2} = \frac{h'_{l,l}}{\varrho}(g'_1 x_{1,2} - g'_1 \beta_1 x_{1,1}) + n_{l,2}. \quad (7)$$

In the fourth channel use, the selected relays precode then forward their respective received signals. Indeed, R$_l$ precodes its received signals by multiplying it by $\frac{(g'_l)^*}{\alpha_l |g'_l|}$. $\alpha_l$ is used to have average transmit power $2mP$ and $\frac{(g'_l)^*}{|g'_l|}$ is used to guarantee that the signals forwarded from the selected relays add constructively at Bob. The received signal by Bob can be written as given by (8), where the variance of $n'_2$ is equal to $\sigma^2(1 + \sum_{1}^{L} \frac{|g'_l|^2}{\alpha_l^2})$. The signals received during the second and fourth channel uses can be written in vector form as

$$\boldsymbol{y} = \begin{bmatrix} y_1 \\ \frac{\varrho}{\sum_{1}^{l} |g'_l|} y_2 \end{bmatrix} = \begin{bmatrix} g'_1 x_{1,1} \\ g'_1 x_{1,2} \end{bmatrix} + \beta_1 \begin{bmatrix} g'_1 x_{1,2} \\ -g'_1 x_{1,1} \end{bmatrix} + \begin{bmatrix} n'_1 \\ \frac{\varrho}{\sum_{1}^{l} |g'_l|} n'_2 \end{bmatrix}. \quad (8)$$

One can easily conclude that the signals other than $x_{1,1}$ and $x_{1,2}$ are confined to (i.e., aligned by) the sub-space formed by the signal vector $(g'_1 x_{1,2}, -g'_1 x_{1,1})^T$ which is orthogonal to $(g'_1 x_{1,1}, g'_1 x_{1,2})^T$. Since $\beta_1$ is aligned by $(g'_1 x_{1,2}, -g'_1 x_{1,1})^T$, it was shown in [14] that the destination can use the received signals $(y_1, y_2)$ to efficiently decode the symbol pair $\{x_{1,1}, x_{1,2}\}$.

In addition to the two first channel use required to deliver the signal $y_1$ to Bob, two channel uses are required to transmit the pair $(x_1, x_2)$, as argued above. Since $y_1$ is used in the decoding process of all symbols, the same applies to the other symbol pairs, that is, each pair requires two channel uses to be delivered to Bob. To elaborate, the symbol pair $(x_3, x_4)$ is precoded and transmitted to the relays in the fifth channel use. To achieve this, Alice uses again the noiseless part of $y_1$, but this time to dissolve $g'_1 \sum_{i=1, i \neq 3}^{2m} x_{1,i} + \sum_{k=2}^{K} g'_k \sum_{i=1}^{2m} x_{k,i}$ in $g'_1 x_{1,4}$, and it also calculates the dissolution factor $\beta_2$. Alice beamforms $g'_1 x_{1,4} - \beta_2 g'_1 x_{1,3}$ in the direction of $\{R_1, R_2, \ldots, R_L\}$. In turn, these relays amplify and forward the received signal during the sixth channel use. Bob uses the signals $(y_1, y_3)$ to decode the second signal pair $\{x_{1,3}, x_{1,4}\}$. Alice, the selected relays and Bob proceed similarly for the remaining signal pairs. In general, during the $(2m(l-1) + 2i - 1)^{th}$ channel use for ($\{i, l\} \in [1, m] \times [1, L]$), the $(m(l-1) + i)^{th}$ symbol pair is precoded then bemformed in the direction of all selected relays. During the $2(m(l-1) + i + 1)^{th}$ channel use, each selected relay amplifies and forwards its respective received signal. The destination then uses $(y_1, y_{m(l-1)+i+1})$ to decode the $(m(l-1) + i)^{th}$ symbol pair.

Based on the above discussion, we conclude that $2(mL+1)$ channel uses are required to transmit $2mL$ symbols, resulting in rate $\frac{2mL}{2(mL+1)} \underset{m \to \infty}{\to} 1$ symbol per channel use. We stress here that ID achieves rate two symbols per channel use for a point-to-point system [14], whereas, the achievable rate for the underlying system is halved since it is a two-hop link and the relays are half-duplex.

As per the ID scheme, for each symbol pair, the remaining signals are nonlinearly precoded in order to be aligned by the intended symbol vector. Since the channel gain vector between the transmitter and any relay (selected or not) is different from the transmitter-relays-destination's channel vector, the remaining signals will not be aligned at these relays. Moreover, during the first channel use, the transmitter communicates the sums of $2m$ symbols $\{\sum_{i=1}^{2m} x_{1,i}, \sum_{i=}^{2m} x_{2,i}, \ldots, \sum_{i=1}^{2m} x_{L,i}\}$ which implies that each symbol is aligned, i.e., received via the same channel, with $2m - 1$ other symbols on the same channel. Therefore, using interference management techniques such as real interference alignement [15]–[17], a relay can only decode the sum of $2m$ symbols but it cannot separate them. The implication here is that any of the selected relays cannot decode symbols individually and hence can not extract any useful information.

## III. ACHIEVABLE SECRECY RATE

In this paper, we make use of the following expression for the achievable secrecy rate [10], [11].

$$R_s = \frac{1}{\tau} \max\left(0, I(\boldsymbol{x}^m; \boldsymbol{y}^m) - \max_{k \in [1, K]} I(\boldsymbol{x}^m; \boldsymbol{z}_k^m)\right), \quad (9)$$

where $\tau$ is the number of channel uses and $\boldsymbol{x}_k^m$ is the channel input. The channel output is denoted by the pair $\{\boldsymbol{z}_k^m, \boldsymbol{y}^m\}$, which represent the received signals by R$_k$ and Bob, respectively. We note that this secrecy rate can be achieved in the sense of strong secrecy by using the channel resolvability-based method to code the message [19].[1]

To analyze (9), we need to analyze the achievable rate on all subchannels, including the Alice-Bob, Alice-relay channels. The latter includes the selected and non-selected relays. As shown above, the symbols are precoded and decoded in pairs. Moreover, the precoding and decoding processes are similar for all symbol pairs. Therefore, without loss of generality, we consider the symbol pair $\{x_{1,1}, x_{1,2}\}$ in our analysis. We then generalize the result to the remaining $mL - 1$ symbol pairs. Based on this result, we provide a lower bound on the achievable secrecy rate at high SNR. These results are used to provide a lower bound on the achievable secrecy diversity order.

---

[1]Precoding and coding in this paper are used to denote two different signal processing stages. In the case of PLS, coding is used to map a binary sequence to a sequence of symbols in order to achieve weak or strong secrecy. Whereas precoding is used to maximize Alice-Bob's mutual information and to degrade Alice-Eve's channel.

### A. The Achievable Rate for the Alice-Bob Channel

The transmitted symbols and the noise components are assumed to be Gaussian. Therefore, the signals $(y_1, y_2)$ tend to be Gaussian and the achievable rate associated to the symbol pair $\{x_{1,1}, x_{1,2}\}$ at the destination is lower bounded as given by the following lemma.

**Lemma 1.** *The achievable rate associated to the symbol pair $\{x_{1,1}, x_{1,2}\}$ on the Alice-Bob channel is lower bounded as*

$$R_{Bob}(x_{1,1}, x_{1,2}) \geqslant \log_2\left(1 + \frac{2mP \sum_{l=1}^{L} |g'_l|^2}{\sigma^2 \left(1 + \sum_{l=1}^{L} \frac{|g_l|^2}{\alpha_l^2}\right)}\right) - 1. \tag{10}$$

*Proof.* See Appendix A. □

It is clear from (10) that the lower bound is independent of the symbol pair index, and it only depends on the selected relays. As such, the same bound applies to all symbol pairs. Recall that the number of channel uses required to transmit the $2mL$ symbols is $2(mL + 1)$. Consequently, the average achievable rate at Bob for large values of $mL$ is given as

$$\begin{aligned}
R_{\text{Bob}} &= \frac{1}{2(mL+1)} \left(\sum_{l=1}^{L} \sum_{i=1}^{m} R(x_{l,2i-1}, x_{l,2i})\right) \\
&\geqslant \frac{1}{2(mL+1)} \sum_{l=1}^{L} \sum_{i=1}^{m} \left[\log_2\left(1 + \frac{2mP \sum_{l=1}^{L} |g'_l|^2}{\sigma^2 \left(1 + \sum_{l=1}^{L} \frac{|g_l|^2}{\alpha_l^2}\right)}\right) - 1\right] \\
&\simeq \frac{1}{2} \log_2\left(1 + \frac{2mP \sum_{l=1}^{L} |g'_l|^2}{\sigma^2 \left(1 + \sum_{l=1}^{L} \frac{|g_l|^2}{\alpha_l^2}\right)}\right) - \frac{1}{2}.
\end{aligned} \tag{11}$$

Note that the transmission of the $mL$ symbols is performed within the same (finite) coherence time, which might appear as a contradiction with the assumption of having large values of $mL$. The assumption of having large values of $mL$ is considered merely to determine the multiplexing gain (pre-log factor). The original expression is $\frac{mL}{2(mL+1)}$. When $mL$ is very large, this expression approaches $\frac{1}{2}$. This is true even if $mL$ is not very large. For example, when $mL = 20$, the expression becomes $\frac{1}{2} - \frac{1}{42} \simeq \frac{1}{2}$. It should be emphasized here that this assumption has nothing to do with the achievable secrecy diversity order, which is the focus of this paper.

In the sequel, we are interested in determining the achievable secrecy diversity order, which is valid at high SNR. In this case, we have

$$\alpha_l^2 = |h'_{l,l}|^2 + \frac{\sigma^2}{2mP} \simeq |h'_{l,l}|^2 \quad \text{and} \quad |g'_l|^2 = \frac{|g_l|^2 |h'_{l,l}|^2}{\alpha_l^2} \simeq |g_l|^2.$$

Consequently, the expression in (11) simplifies at high SNR as

$$R_{\text{Bob}} \geqslant \frac{1}{2} \log_2 \left( 1 + \frac{2mP \sum_{l=1}^{L} |g_l|^2}{\sigma^2 \left( 1 + \sum_{l=1}^{L} \frac{|g_l|^2}{|h'_{l,l}|^2} \right)} \right) - \frac{1}{2}. \quad (12)$$

*B. The Achievable rate for the Alice-Selected Relay channels*

In this subsection, we first derive the achievable rate associated to a selected relay considering the symbol pair $\{x_{1,1}, x_{1,2}\}$. We then generalize it to the remaining symbol pairs. Recall that the received signals by $R_l$ during the first and third channel uses can be written as

$$\begin{bmatrix} z_{l,1} \\ z_{l,2} \end{bmatrix} = \begin{bmatrix} h'_{l,l} \sum_{i=1}^{2m} x_{l,i} \\ \frac{h'_{l,l}}{\sum_{l=1}^{L} |g'_l|^2} (g'_1 x_{1,2} - g'_1 \beta_1 x_{1,1}) \end{bmatrix} + \begin{bmatrix} n_{l,1} \\ n_{l,2} \end{bmatrix}, \quad (13)$$

where $n_{l,1}$ and $n_{l,2}$ are both AWGN with zero mean and variance $\sigma^2$.

Since $\beta_1$ in (11) is a nonlinear combination of signals and the vector channel $\{g'_1, g'_2, \ldots, g'_L\}$, $z_{l,1}$ cannot be written as $g'_1(x_{1,1} + \beta_1 x_{1,2}) + n_{1,1}$. Hence, the remaining signals are not aligned by the intended one as in (8). Although Eve collocated with $R_l$ has the Alice-Bob's CSI, it cannot proceed similar to Bob to decode $x_{1,1}$ and $x_{1,2}$ because the remaining symbols are not aligned by the intended one at the relays. Moreover, each symbol is aligned with $2m - 1$ symbols on the same channel in $z_{1,1}$, i.e., received over the same channel as $2m - 1$ symbols. Hence, Eve cannot use real interference alignment [15]–[17] to separate the signals. In $z_{1,2}$, $\beta_1$ is unknown and Eve cannot use this signal to extract any of the two symbols. This heuristic interpretation suggests, from an information theoretic perspective, that the proposed technique achieves very low rate at Eve, which is proved in the following Lemma.

**Lemma 2.** *At high SNR, the achievable rate associated to the symbol pair $\{x_{1,1}, x_{1,2}\}$ on the Alice $-R_l$ ($l \in [1, L]$) channel is given by:*

$$R_{R_l}(x_{1,1}, x_{1,2}) \simeq \log_2 \left( 1 + \frac{|g_l|^2}{\sum_{\substack{j=1 \\ j \neq l}}^{L} |g_j|^2} \right). \quad (14)$$

*Proof.* See Appendix B. □

The expression of the achievable rate in (14) is independent of the index of the transmitted pair of symbols. It depends only on the considered relays. Therefore, this expression is valid for all other symbol pairs. Consequently, the achievable rate per channel use at high SNR at a given selected relay is upper bounded as

$$\begin{aligned} R_{R_l} &\simeq \frac{m}{2(m+1)} \log_2 \left( 1 + \frac{|g_l|^2}{\sum_{j=1, j \neq l} |g_j|^2} \right) \\ &\simeq \frac{1}{2} \log_2 \left( 1 + \frac{|g_l|^2}{\sum_{j=1, j \neq l} |g_j|^2} \right) \\ &\leqslant \frac{1}{2} \max_{l \in [1,L]} \log_2 \left( 1 + \frac{|g_l|^2}{\sum_{\substack{j=1 \\ j \neq l}}^{L} |g_j|^2} \right). \end{aligned} \quad (15)$$

*C. The Achievable Rate for the Alice-Non Selected Relays Channels*

Let us now consider a non selected relay belonging to the set $\{R_{L+1}, R_{L+2}, \ldots, R_K\}$. We adopt the same strategy as in the previous section where we first consider the symbol pair $\{x_{1,1}, x_{1,2}\}$. A non selected relay $R_l$ $l \in [L+1, K]$ can take advantage of the signal transmitted by Alice during the first and third channel uses, in order to decode $\{x_{1,1}, x_{1,2}\}$, which can be written in vector form as

$$\begin{bmatrix} z_{l,1} \\ z_{l,2} \end{bmatrix} = \begin{bmatrix} \sum_{l=1}^{L} h'_{l,j} \sum_{i=1}^{2m} x_{l,i} \\ \frac{\sum_{l=1}^{L} h'_{l,j}}{\varrho} (g'_1 x_{1,2} - \beta_1 g'_1 x_{l,1}) \end{bmatrix} + \begin{bmatrix} n_{j,1} \\ n_{j,2} \end{bmatrix}, \quad (16)$$

where $n_{l,1}$ and $n_{l,2}$ are both AWGN with zero mean and variance $\sigma^2$. In the second line of (16), we use $h'_{l,j}$ to denote $\frac{h_l b_j^T}{\|b_j\|}$. We hereafter use $h'_l$ to denote the vector $(h'_{l,1}, h'_{l,2}, \ldots, h'_{l,L})^T$.

**Lemma 3.** *At high SNR, the achievable rate associated to the symbol pair $(x_{1,1}, x_{1,2})$ on the Alice $-R_l$ ($l \in [L+1, K]$) channel, does not scale with power for almost all channel realizations and it can be written as*

$$R_{R_l}(x_{1,1}, x_{1,2}) \simeq \log_2 \left( \frac{\|h'_l\|^2 \|g\|^2}{\|h'_l\|^2 \|g\|^2 - \langle h'_l, g \rangle^2} \right). \quad (17)$$

*Proof.* See Appendix C. □

Given that the expression in (17) does not depend on the symbol pair index, the achievable rate at high SNR at a given non-selected relay is given as

$$\begin{aligned} R_{R_l} &\simeq \frac{m}{2(m+1)} \log_2 \left( \frac{\|h'_l\|^2 \|g\|^2}{\|h'_l\|^2 \|g\|^2 - \langle h'_l, g \rangle^2} \right) \\ &\simeq \frac{1}{2} \log_2 \left( \frac{\|h'_l\|^2 \|g\|^2}{\|h'_l\|^2 \|g\|^2 - \langle h'_l, g \rangle^2} \right). \end{aligned} \quad (18)$$

Now that we have obtained expressions for the achievable rate on all sub-channels, we proceed in the next section where we make use of these results to determine the achievable secrecy diversity order.

## IV. ACHIEVABLE SECRECY DIVERSITY ORDER

Given that the underlying channel is Rayleigh fading, there is a relationship between the outage probability and the achievable secrecy rate. Specifically, $P_{out}(\text{SNR}, \gamma) = P_r [R_s(\text{SNR}) < \gamma]$ where SNR is defined as $\text{SNR} \triangleq \frac{2mP}{\sigma^2}$ and

$\gamma$ is a secrecy rate threshold [20]. Consequently, the achievable secrecy diversity order is defined as

$$d \triangleq \lim_{\text{SNR} \to \infty} \frac{-\log P_{out}(\text{SNR}, \gamma)}{\log \text{SNR}}.$$

In Section III, we provided a lower bound on the achievable rate on the Alice-Bob channel. Moreover, we provided an upper bound on the achievable rate on the Alice-relay channel (selected and non selected). These give a lower bound on the achievable secrecy rate. Since the diversity order is by definition computed asymptotically at high SNR, we make use of the secrecy rate lower bound to provide an upper bound on the outage probability at high SNR, which is used later to provide a lower bound on the achievable diversity order. The secrecy rate is given (9) and can be lower bounded at high SNR as

$$\begin{aligned} R_s &= \max\left(0, R_{\text{Bob}} - \max_{l \in [1,K]} R_{\text{R}_l}\right) \\ &\geqslant \max\Bigg\{0, \frac{1}{2}\log_2\left(1 + \frac{2mP \sum_{l=1}^{L} |g_l|^2}{\sigma^2 \left(1 + \sum_{l=1}^{L} \frac{|g_l|^2}{|h'_{l,l}|^2}\right)}\right) - \frac{1}{2} \\ &\quad - \frac{1}{2}\max\Bigg[\max_{l \in [1,L]} \log_2\left(1 + \frac{|g_l|^2}{\sum_{\substack{j=1 \\ j \neq l}}^{L} |g_j|^2}\right), \\ &\quad \max_{l \in [L+1,K]} \log_2\left(\frac{\|\boldsymbol{h}'_l\|^2 \|\boldsymbol{g}\|^2}{\|\boldsymbol{h}'_l\|^2 \|\boldsymbol{g}\|^2 - \langle \boldsymbol{h}'_l, \boldsymbol{g} \rangle^2}\right)\Bigg]\Bigg\}. \end{aligned} \tag{19}$$

On the other hand, the outage probability can be written as

$$\begin{aligned} P_{out}(\text{SNR}, \gamma) &= P_r\left[\max\left(0, R_{\text{Bob}} - \max_{l \in [1,K]} R_{\text{R}_l}\right) < \gamma\right] \\ &= \underbrace{P_r[0 < \gamma]}_{=1} P_r\left[\left(R_{\text{Bob}} - \max_{l \in [1,K]} R_{\text{R}_l}\right) < \gamma\right] \\ &= P_r\left[\left(R_{\text{Bob}} - \max_{l \in [1,K]} R_{\text{R}_l}\right) < \gamma\right]. \end{aligned} \tag{20}$$

Next, we show that for a given positive real constant $\epsilon \ll 1$, which can be as small as desired, there exists a $\text{SNR}_{th}$ such that $\max_{l \in [1,K]} R_l$ can be upper bounded by $\frac{1}{2}\log_2(\text{SNR}^\epsilon)$ $\forall \text{SNR} \geqslant \text{SNR}_{th}$, i.e., $\lim_{\text{SNR} \to \infty} P_r\left[\max_{l \in [1,K]} R_l < \frac{1}{2}\log_2(\text{SNR}^\epsilon)\right] = 1$. This bound is used later to provide an upper bound on the outage probability.

In (21) on the next page, we provide an expression for the outage probability as a function of $P_{out1}(\text{SNR}^\epsilon) = P_r\left[\max_{l \in [1,K]} R_l < \frac{1}{2}\log_2(\text{SNR}^\epsilon)\right]$.

**Lemma 4.** $P_{out1}(SNR^\epsilon) \underset{SNR \to \infty}{\to} 1$

*Proof.* See Appendix D. $\square$

Invoking the result in Lemma 4, the upper bound on the outage probability given in (21) becomes

$$\begin{aligned} &P_{out}(\gamma, \text{SNR}) \\ &\leqslant P_r\left[R_{\text{Bob}} - \frac{1}{2}\log_2(\text{SNR}^\epsilon) < \gamma \Big| \max_{l \in [1,K]} R_l < \frac{1}{2}\log_2(\text{SNR}^\epsilon)\right]. \end{aligned} \tag{22}$$

**Lemma 5.** *At high SNR and for a given small constant $\epsilon$, the outage probability can be upper bounded as*

$$P_{out}(\gamma, SNR) \leqslant P_r\left[\sum_{\substack{l=1 \\ l \neq l_m}}^{L} |g_l|^2 < \frac{2^{2\gamma+1}}{\left(SNR^{1-\epsilon} - 2^{2\gamma+2}SNR^\epsilon\right)}\right], \tag{23}$$

*where* $l_m = \arg\max_{l \in [1,L]}(|g_l|^2)$.

*Proof.* See Appendix E. $\square$

Since the components of channel vector $\boldsymbol{g}$ follow the Rayleigh distribution, $\sum_{l=1, l \neq l_m}^{L} |g_l|^2$ follows a Chi-square distribution with order $2(L - 1)$. From [20] and (19), we express the achievable secrecy diversity order as

$$\begin{aligned} d &= \lim_{\text{SNR} \to \infty} \frac{-\log P_r[R_s \leqslant \gamma]}{\log \text{SNR}} \\ &\geqslant \lim_{\text{SNR} \to \infty} \frac{-\log P_r\left[\sum_{\substack{l=1 \\ l \neq l_m}}^{L} \left(\text{SNR}^{1-\epsilon} - 2^{2\gamma+2}\text{SNR}^\epsilon\right)|g_l|^2 < 2^{2\gamma+1}\right]}{\log \text{SNR}} \\ &\simeq \lim_{\text{SNR} \to \infty} \frac{-\log P_r\left[\sum_{\substack{l=1 \\ l \neq l_m}}^{L} \text{SNR}^{1-\epsilon}|g_l|^2 < 2^{2\gamma+1}\right]}{\log \text{SNR}} \\ &= (L-1)(1-\epsilon). \end{aligned} \tag{24}$$

Since $\epsilon$ can be as small as desired, the proposed technique can therefore achieve a secrecy diversity order equal to the number of selected relays $L$. The only condition imposed on the number of selected relays is $2 \leqslant L \leqslant \min(N, K)$, suggesting that Alice can select up to $(\min(N, K) - 1)$ relays and this yields a secrecy diversity order of up to $\min(N, K) - 1$.

## V. NUMERICAL RESULTS

### A. System and Simulation Setup

In this section, we provide simulation and numerical results to validate the analytical results obtained in the previous sections. It is assumed that Alice is equipped with four antennas, and there are 10 relays. The number of selected relays varies from two to four. Bob is assumed to be equipped with a single antenna. All channels are Rayleigh distributed with variance one. We assume that each relay can provide an average SNR of $\frac{2mP}{\sigma^2}$ which is independent from the number of the selected relays. The SNR considering in simulation results is the SNR per relay. This implies that the total used transmit power increases as the number of the selected relays increase. However, increasing the total transmit power does not affect the secrecy diversity order.

$$\begin{aligned}
&P_{out}(\text{SNR}, \gamma) \\
&= P_r\left[R_{\text{Bob}} - \max_{l \in [1,K]} R_{R_l} < \gamma\right] \\
&= (1 - P_{out1}(\text{SNR}^\epsilon))P_r\left[R_{\text{Bob}} - \max_{l \in [1,K]} R_{R_l} < \gamma | \max_{l \in [1,K]} R_l \geq \frac{1}{2}\log_2(\text{SNR}^\epsilon)\right] + P_{out1}(\text{SNR}^\epsilon)P_r\left[R_{\text{Bob}} - \max_{l \in [1,K]} R_{R_l} < \gamma | \max_{l \in [1,K]} R_l < \frac{1}{2}\log_2(\text{SNR}^\epsilon)\right] \\
&\leq (1 - P_{out1}(\text{SNR}^\epsilon))P_r\left[R_{\text{Bob}} - \max_{l \in [1,K]} R_{R_l} < \gamma | \max_{l \in [1,K]} R_l \geq \frac{1}{2}\log_2(\text{SNR}^\epsilon)\right] + P_{out1}(\text{SNR}^\epsilon)P_r\left[R_{\text{Bob}} - \frac{1}{2}\log_2(\text{SNR}^\epsilon) < \gamma | \max_{l \in [1,K]} R_l < \frac{1}{2}\log_2(\text{SNR}^\epsilon)\right] \\
&\hspace{15cm} (21)
\end{aligned}$$

## B. Achievable secrecy diversity order

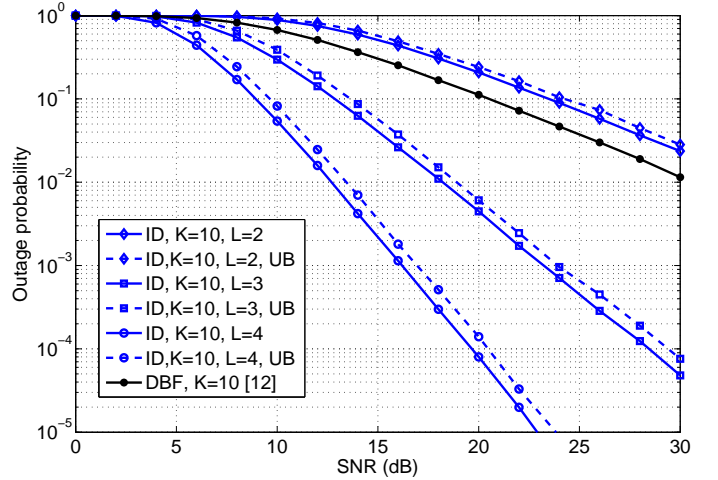

Fig. 2: Outage probability at the destination versus SNR(dB).

We evaluate the achievable secrecy diversity order as a function of the SNR in dB for the above system setup. We plot the outage probability in Fig. 2 where the secrecy rate threshold is set to $\gamma = 1$. In the figure, we compare the exact outage probability with the corresponding upper bound. The exact expression for the achievable rate on the Alice-Bob channel is given by (25), where the exact expressions of the denominator and nominator are given by (26) and (27), respectively. The other exact expressions of the achievable rate on the other channels can be found in Appendices B and C. The results corresponding to the upper bound on the outage probability (UB) are obtained by considering the asymptotic lower bound on the secrecy rate given in (19). It is clear from the figure that the outage probability slope, for the proposed technique, is equal to $L-1$ and hence it matches the theoretical lower bound. The asymptotic lower bound is close to the exact outage probability which proves the tightness of the provided secrecy rate lower bound.

We also compare in the same figure the performance of the proposed scheme with the distributed beamforming (DBF) scheme proposed in [12] which provides, to the best of our knowledge, the best performance in terms of the achievable diversity order. As shown in the figure, the proposed technique outperforms the DBF technique when $L > 2$. Both schemes have the same secrecy diversity order when $L = 2$, however the DBF outperforms the proposed scheme for this particular case. The reason is that the results are plotted against the SNR in dB per relay, and this makes it more advantageous for the DBF since all relays are used for relaying, whereas only two relays are used in the proposed scheme.

## C. Achievable secrecy rate

Fig. 3 depicts the achievable secrecy rate as a function of SNR for different numbers of untrusted relays, namely, $K = 4, 6, 8, 10, 12$. The exact expression of the secrecy rate is considered in this figure. We set $N = L = 3$. The figure shows that the secrecy rate increases with the SNR, which confirms that the rates associated to the untrusted relays do not scale with power. The figure shows also that the secrecy rate decreases as the number of untrusted relays increases. In fact, the secrecy rate is equal to the difference between the achievable rate at Bob and the maximum of the achievable rates corresponding to the untrusted relays, which increases as $K$ increases. However, the loss in the secrecy rate is not considerable as the number of untrusted relays increases. The secrecy rate decreases by approximately a half bit as the number of untrusted relays goes from $4$ to $10$. This proves the robustness of the proposed technique to the number of untrusted relays.

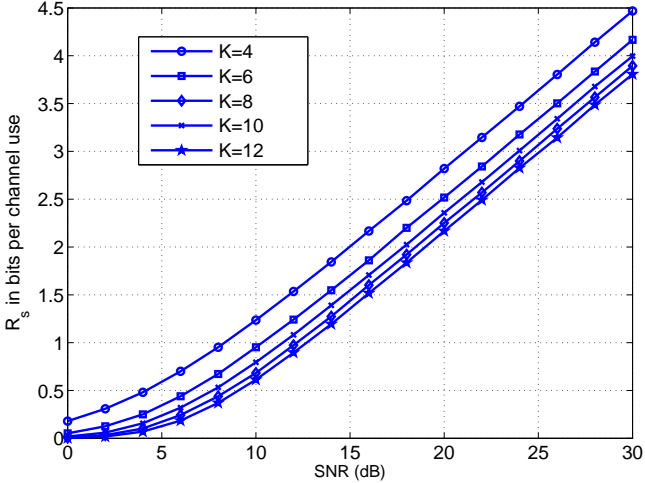

Fig. 3: Achievable secrecy rate versus SNR(dB).

## VI. Conclusion

We studied in this paper the achievable secrecy diversity order of cooperative networks with untrusted relays. We considered a two-hop network comprising an $N$-antenna Alice, $K$ single-antenna relays and a single antenna Bob. We proposed a nonlinear PLS technique, based on a previously proposed interference alignment scheme, that ensures secure communication between Alice and Bob via the untrusted relays. The proposed scheme was analyzed in terms the achievable secrecy rate and secrecy diversity order. It was shown that a secrecy diversity order of up to $(\min(N, K) - 1)$ is achievable. This is an important result because it is contrary to what has been published so far on this subject. In particular, it has been shown that the achievable secrecy diversity order is one for the case when $K$ may higher than $N$. Furthermore, the proposed method does not use artificial noise which is deemed the only option to secure communications for the adopted system model. The achieved performance is based on random relay selection. We believe that a proper relay selection method can dramatically enhance the achievable diversity order.

## Appendix A

Since the transmitted symbols and the noise components are Gaussian, the signals $(y_1, y_2)$ tend to be Gaussian and the achievable rate associated to the symbol pair $\{x_{1,1}, x_{1,2}\}$ at the destination is written as

$$\begin{aligned} R_{\text{Bob}}(x_{1,1}, x_{1,2}) &= I(x_{1,1}, x_{1,2}; y_1, y_2) \\ &= H(y_1, y_2) - H(y_1, y_2 | x_{1,1}, x_{1,2}) \\ &= \log_2 \left( \frac{|C(y_1, y_2)|}{E[|C(y_1, y_2 | x_{1,1}, x_{1,2})|]} \right), \end{aligned} \quad (25)$$

where $C(y_1, y_2)$ and $C(y_1, y_2 | x_{1,1}, x_{1,2})$ are the covariances of $(y_1, y_2)$ and $(y_1, y_2)$ given $(x_{1,1}, x_{1,2})$, respectively. Their explicit formulas are given in (26) and (27) on the next page, respectively. The first inequality in (27) comes from the fact that $\frac{(\sum_{l=1}^{L} |g'_l|)^2}{\varrho^2} = \frac{(\sum_{l=1}^{L} |g'_l|)^2}{\sum_{l=1}^{L} |g'_l|^2} \geqslant 1$ and hence $1 + \frac{(\sum_{l=1}^{L} |g'_l|)^2}{\varrho^2} \leqslant 2 \frac{(\sum_{l=1}^{L} |g'_l|)^2}{\varrho^2}$.

Substituting (27) in (25), we obtain a lower bounded on the achievable rate as follows.

$$\begin{aligned} R_{\text{Bob}}(x_{1,1}, x_{1,2}) &\geqslant \log_2 \left( 1 + \frac{2mP\varrho^2}{\sigma^2 \left(1 + \sum_{l=1}^{L} \frac{|g_l|^2}{\alpha_l^2}\right)} \right) + \log_2 \left(\frac{1}{2}\right) \\ &= \log_2 \left( 1 + \frac{2mP \sum_{l=1}^{L} |g'_l|^2}{\sigma^2 \left(1 + \sum_{l=1}^{L} \frac{|g_l|^2}{\alpha_l^2}\right)} \right) + \log_2 \left(\frac{1}{2}\right) \\ &= \log_2 \left( 1 + \frac{2mP \sum_{l=1}^{L} |g'_l|^2}{\sigma^2 \left(1 + \sum_{l=1}^{L} \frac{|g_l|^2}{\alpha_l^2}\right)} \right) - 1, \end{aligned} \quad (28)$$

which proves (10).

## Appendix B

The rate associated to $(x_{1,1}, x_{1,2})$ given $(z_{1,1}, z_{1,2})$ is:

$$\begin{aligned} R_{\text{R}_l}(x_{1,1}, x_{1,2}) &= I(x_{1,1}, x_{1,2}; z_{l,1}, z_{l,2}) \\ &= H(z_{l,1}, z_{l,2}) - H(z_{l,1}, z_{l,2} | x_{1,1}, x_{1,2}) \\ &= \log_2 \left( \frac{|C(z_{l,1}, z_{l,2})|}{E[|C(z_{l,1}, z_{l,2} | x_{1,1}, x_{1,2})|]} \right). \end{aligned} \quad (29)$$

Since only $\text{R}_1$ receives a signal depending on $\{x_{1,1}, x_{1,2}\}$ in the first channel use, the expression of $E[|C(z_{l,1}, z_{l,2} | x_{1,1}, x_{1,2})|]$ differs slightly from the remaining covariance matrices. Explicit expressions of the covariance matrices are given in (30), (31) and (32), where the expression in (32) is valid for all $l \in [2, L]$. At large values of $m$, we have $m - 1 \simeq m$ and hence the covariance in (31) can be written as (32) by replacing $l$ by one. Next, we consider the case of large values of $m$ and we thus use the general expression provided in (32). The expressions of the covariance matrix in (30) and (32) can be written in the form $(PC_1 + \sigma^2)(PC_2 + \sigma^2)$ and $\sigma^2(PC_3 + \sigma^2) + P^2 C_4$, respectively, where $\{C_1, C_2, C_3, C_4\}$ are constants at high SNR. They depend only on the index of the considered relay. The achievable rate can be written as

$$R_{\text{R}_l}(x_{1,1}, x_{1,2}) = \log_2 \left( \frac{(PC_1 + \sigma^2)(PC_2 + \sigma^2)}{\sigma^2(PC_3 + \sigma^2) + P^2 C_4} \right). \quad (33)$$

We observe from (33) that, at high SNR, the denominator scales with $P^2$. Given that the denominator also scales with $P^2$, the achievable rate becomes a constant at high SNR. That is,

$$\begin{aligned} R_{\text{R}_l}(x_{1,1}, x_{1,2}) &\simeq \log_2 \left( \frac{C_1 C_2}{C_4} \right) \\ &\simeq \log_2 \left( \frac{\sum_{j=1}^{L} |g_j|^2}{\sum_{j=1, j \neq l}^{L} |g_j|^2} \right) \\ &\simeq \log_2 \left( 1 + \frac{|g_l|^2}{\sum_{j=1, j \neq l}^{L} |g_j|^2} \right), \end{aligned} \quad (34)$$

which proves (14).

$$|C(y_1, y_2)| = \begin{vmatrix} E[|y_1|^2] & E[y_1(y_2)^*] \\ E[(y_1)^* y_2] & E[|y_2|^2] \end{vmatrix}$$
$$= \begin{vmatrix} 2mP\varrho^2 + \sigma^2\left(1 + \sum_{l=1}^L \frac{|g_l|^2}{\alpha_l^2}\right) & 0 \\ 0 & \frac{(\sum_{l=1}^L |g_l'|)^2}{\varrho^2} 2mP\varrho^2 + \sigma^2\left(1 + \sum_{l=1}^L \frac{|g_l|^2}{\alpha_l^2}\right) \end{vmatrix} \quad (26)$$
$$= \left(2mP\varrho^2 + \sigma^2\left(1 + \sum_{l=1}^L \frac{|g_l|^2}{\alpha_l^2}\right)\right) \underbrace{\left(2mP(\sum_{l=1}^L |g_l'|)^2 + \sigma^2\left(1 + \sum_{l=1}^L \frac{|g_l|^2}{\alpha_l^2}\right)\right)}_{term1}.$$

$$E[|C(y_1, y_2|x_{1,1}, x_{1,2})|] = E\left[\begin{vmatrix} E[|y_1|^2|x_{1,1}, x_{1,2}] & E[y_1 y_2^*|x_{1,1}, x_{1,2}] \\ E[y_1^* y_2|x_{1,1}, x_{1,2}] & E[|y_2|^2|x_{1,1}, x_{1,2}] \end{vmatrix}\right]$$
$$= E\left[\begin{vmatrix} 2P\left((m-1)|g_1'|^2 + m\sum_{l=2}^L |g_l'|^2\right) + \sigma^2\left(1 + \sum_{l=1}^L \frac{|g_l|^2}{\alpha_l^2}\right) & -2P\sqrt{\frac{(\sum_{l=1}^L |g_l'|)^2}{\varrho^2}}\left((m-1)|g_1'|^2 + m\sum_{l=2}^L |g_l'|^2\right)\left(\frac{x_{1,1}}{x_{1,2}}\right)^* \\ -2P\sqrt{\frac{(\sum_{l=1}^L |g_l'|)^2}{\varrho^2}}\left((m-1)P|g_1'|^2 + m\sum_{l=2}^L |g_l'|^2\right)\frac{x_{1,1}}{x_{1,2}} & 2P\frac{(\sum_{l=1}^L |g_l'|)^2}{\varrho^2}\left((m-1)P|g_1'|^2 + m\sum_{l=2}^l |g_l'|^2\right)\frac{x_{1,1}^2}{x_{1,2}^2} + \sigma^2\left(1 + \sum_{l=1}^L \frac{|g_l|^2}{\alpha_l^2}\right) \end{vmatrix}\right]$$
$$= \left(2(m-1)P|g_1'|^2 + 2mP\sum_{l=2}^L |g_l'|^2 + \sigma^2\left(1 + \sum_{l=1}^L \frac{|g_l|^2}{\alpha_l^2}\right)\right)\left(\frac{(\sum_{l=1}^L |g_l'|)^2}{\varrho^2}\left(2(m-1)P|g_1'|^2 + mP\sum_{l=2}^L |g_l'|^2\right) E\left[\frac{x_{1,1}^2}{x_{1,2}^2}\right]\right.$$
$$\left. + \sigma^2\left(1 + \sum_{l=1}^L \frac{|g_l|^2}{\alpha_l^2}\right)\right) - \frac{(\sum_{l=1}^L |g_l'|)^2}{\varrho^2}\left(2(m-1)P|g_1'|^2 + 2mP\sum_{l=2}^L |g_l'|^2\right)^2 E\left[\frac{x_{1,1}^2}{x_{1,2}^2}\right]$$
$$= \sigma^2\left(1 + \sum_{l=1}^L \frac{|g_l|^2}{\alpha_l^2}\right)\left[2P\left(1 + \frac{(\sum_{l=1}^L |g_l'|)^2}{\varrho^2}\right)\left((m-1)|g_1'|^2 + m\sum_{l=2}^L |g_l'|^2\right) + \sigma^2\left(1 + \sum_{l=1}^L \frac{|g_l|^2}{\alpha_l^2}\right)\right]$$
$$\leqslant \sigma^2\left(1 + \sum_{l=1}^L \frac{|g_l|^2}{\alpha_l^2}\right)\left[4P\frac{(\sum_{l=1}^L |g_l'|)^2}{\varrho^2}\left((m-1)|g_1'|^2 + m\sum_{l=2}^L |g_l'|^2\right) + \sigma^2\left(1 + \sum_{l=1}^L \frac{|g_l|^2}{\alpha_l^2}\right)\right]$$
$$\leqslant \sigma^2\left(1 + \sum_{l=1}^L \frac{|g_l|^2}{\alpha_l^2}\right)\left[4P\frac{(\sum_{l=1}^L |g_l'|)^2}{\varrho^2}\left(m|g_1'|^2 + m\sum_{l=2}^L |g_l'|^2\right) + 2\sigma^2\left(1 + \sum_{l=1}^L \frac{|g_l|^2}{\alpha_l^2}\right)\right]$$
$$= \sigma^2\left(1 + \sum_{l=1}^L \frac{|g_l|^2}{\alpha_l^2}\right)\left[\underbrace{4mP(\sum_{l=1}^L |g_l'|)^2 + 2\sigma^2\left(1 + \sum_{l=1}^L \frac{|g_l|^2}{\alpha_l^2}\right)}_{2\times term1}\right].$$
(27)

$$|C(z_{l,1}, z_{l,2})| = \begin{vmatrix} 2mP|h_{l,l}'|^2 + \sigma^2 & 0 \\ 0 & \frac{2mP|h_{l,l}'|^2}{\varrho^2}\varrho^2 + \sigma^2 \end{vmatrix}$$
$$= \left(2mP|h_{l,l}'|^2 + \sigma^2\right)\left(2mP|h_{l,l}'|^2 + \sigma^2\right). \quad (30)$$

$$E[|C(z_{1,1}, z_{1,2}|x_{1,1}, x_{1,2})|] = E\left[\begin{vmatrix} 2P(m-1)|h_{1,1}'|^2 + \sigma^2 & -\left(\frac{x_{1,1}}{x_{1,2}}\right)^* 2P(m-1)\frac{|h_{1,1}'|^2}{\varrho}(g_1')^* \\ -\frac{x_{1,1}}{x_{1,2}} 2P(m-1)\frac{|h_{1,1}'|^2}{\varrho} g_1' & \frac{|x_{1,1}|^2}{|x_{1,2}|^2} 2P\frac{|h_{1,1}'|^2}{\varrho^2}\left((m-1)|g_1'|^2 + m\sum_{j=2}^L |g_j'|^2\right) + \sigma^2 \end{vmatrix}\right]$$
$$= \sigma^2\left(2(m-1)P|h_{1,1}'|^2 + E\left[\frac{|x_{1,1}|^2}{|x_{1,2}|^2}\right]2P\frac{|h_{1,1}'|^2}{\varrho^2}\left((m-1)|g_1'|^2 + m\sum_{j=2}^L |g_j'|^2\right) + \sigma^2\right) \quad (31)$$
$$+ E\left[\frac{|x_{1,1}|^2}{|x_{1,2}|^2}\right] 4P^2 m(m-1)|h_{1,1}'|^4 \frac{\sum_{j=2}^L |g_j'|^2}{\sum_{j=1}^L |g_j'|^2}.$$

$$E[|C(z_{l,1}, z_{l,2}|x_{1,1}, x_{1,2})|] = E\left[\left|\begin{array}{cc} 2Pm|h'_{l,l}|^2 + \sigma^2 & -\left(\frac{x_{1,1}}{x_{1,2}}\right)^* 2Pm\frac{|h'_{l,l}|^2}{\varrho}(g'_l)^* \\ -\frac{x_{1,1}}{x_{1,2}}2Pm\frac{|h'_{l,l}|^2}{\varrho}g'_l & \frac{|x_{1,1}|^2}{|x_{1,2}|^2}2P\frac{|h'_{l,l}|^2}{\varrho^2}\left((m-1)|g'_1|^2 + m\sum_{j=2}^L |g'_j|^2\right) + \sigma^2 \end{array}\right|\right]$$

$$= \sigma^2\left(2mP|h'_{l,l}|^2 + E\left[\frac{|x_{1,1}|^2}{|x_{1,2}|^2}\right]2P\frac{|h'_{l,l}|^2}{\varrho^2}\left((m-1)|g'_1|^2 + m\sum_{j=2}^L |g'_j|^2\right) + \sigma^2\right) \quad (32)$$

$$+ E\left[\frac{|x_{1,1}|^2}{|x_{1,2}|^2}\right]4P^2|h'_{l,l}|^4\left(m^2\frac{\sum_{j=2,j\neq l}^L |g'_j|^2}{\sum_{j=1}^L |g'_j|^2} + m(m-1)\frac{|g'_1|^2}{\sum_{j=1}^L |g'_j|^2}\right).$$

## APPENDIX C

The rate associated to $(x_{1,1}, x_{1,2})$ given $(z_{l,1}, z_{l,2})$ is expressed as

$$I(x_{1,1}, x_{1,2}; z_{l,1}, z_{l,2}) = \log_2\left(\frac{|C(z_{l,1}, z_{l,2})|}{E\left[|C(z_{l,1}, z_{l,2}|x_{1,1}, x_{1,2})|\right]}\right). \quad (35)$$

Explicit expressions of the covariance matrices are given in (36) and (37) on the next page. These expressions can be written in the form $(PC_5+\sigma^2)(PC_6+\sigma^2)$ and $\sigma^2(PC_7+\sigma^2)+P^2C_8$, respectively, where $\{C_5, C_6, C_7, C_8\}$ are constants at high transmit power. The achievable rate can be written:

$$R_{R_l}(x_{1,1}, x_{1,2}) = \log_2\left(\frac{(PC_5 + \sigma^2)(PC_6 + \sigma^2)}{\sigma^2(PC_7 + \sigma^2) + P^2C_8}\right). \quad (38)$$

At high SNR and when $C_8 \neq 0$, we observe from (38) that the denominator scales with $P^2$. Given that the denominator also scales with $P^2$, the achievable rate becomes a constant at high SNR. The achievable rate scales with the transmit power if and only if $C_8 = 0$. By Cauchy Schwartz inequality, $C_8$ is zero if and only if $(\sqrt{m-1}h'_{l,1}, \sqrt{m}h'_{l,2}, \ldots, \sqrt{m}h'_{l,L})$ is parallel to $(\sqrt{m-1}g'_1, \sqrt{m}g'_2, \ldots, \sqrt{m}g'_L)$, i.e., $(h'_{l,1}, h'_{l,2}, \ldots, h'_{l,L})$ is parallel to $(g'_1, g'_2, \ldots, g'_L)$.

The two channel vectors $(h'_{l,1}, h'_{l,2}, \ldots, h'_{l,L})$ and $(g'_1, g'_2, \ldots, g'_L)$ are independent. The scenario when these two vectors are parallel is practically impossible and thus $R_{R_l}(x_{1,1}, x_{1,2})$ converges to a constant, i.e., does not scale with $P$ at high SNR for almost all channel realizations. Therefore, for almost all channel realizations, the achievable rate, at high SNR and for high values of $m$, can be written as

$$R_{R_l}(x_{1,1}, x_{1,2}) \simeq \log_2\left(\frac{C_5 C_6}{C_8}\right) \quad (39)$$
$$\simeq \log_2\left(\frac{\|\boldsymbol{h}_l\|^2\|\boldsymbol{g}\|^2}{\|\boldsymbol{h}_l\|^2\|\boldsymbol{g}\|^2 - \langle\boldsymbol{h}_l, \boldsymbol{g}\rangle^2}\right),$$

which proves (17)

## APPENDIX D

Note that $\max_{l\in[1,K]} R_l \leq \frac{1}{2}\log_2(\text{SNR}^\epsilon)$ implies that the rate on Alice-relay channel (selected or non selected) is less than $\frac{1}{2}\log_2(\text{SNR}^\epsilon)$. This is equivalent to

$$\begin{cases} \frac{|g_l|^2}{\sum_{\substack{j=1\\j\neq l}}^L |g_j|^2} < \text{SNR}^\epsilon - 1, & \forall l \in [1, L] \\ \frac{\|\boldsymbol{h}'_l\|^2\|\boldsymbol{g}^l\|^2}{\|\boldsymbol{h}'_l\|^2\|\boldsymbol{g}\|^2 - \langle\boldsymbol{h}'_l, \boldsymbol{g}\rangle^2} < \text{SNR}^\epsilon & \forall l \in [L+1, K] \end{cases} \quad (40)$$

To get $P_{out1}(\text{SNR}^\epsilon) \underset{\text{SNR}\to\infty}{\to} 1$, the probability that the events in (40) occur should approach one as SNR tends to infinity. Now we have

$$P_r\left[\frac{|g_l|^2}{\sum_{\substack{j=1\\j\neq l}}^L |g_j|^2} < \text{SNR}^\epsilon - 1, \forall l \in [1, L]\right]$$

$$= P_r\left[\max_{l\in[1,L]}\frac{|g_l|^2}{\sum_{\substack{j=1\\j\neq l}}^L |g_j|^2} < \text{SNR}^\epsilon - 1\right] \quad (41)$$

$$= P_r\left[|g_{l_m}|^2 < (\text{SNR}^\epsilon - 1)\sum_{\substack{j=1\\j\neq l_m}}^L |g_j|^2\right],$$

where $l_m = \arg\max_{l\in[1,L]}(|g_l|^2)$ is the index of the maximum channel gain among the vector $(|g_1|^2, |g_2|^2, \ldots, |g_L|^2)$. The term $(\text{SNR}^\epsilon - 1)\sum_{\substack{j=1\\j\neq l_m}}^L |g_j|^2$ increases as SNR increases. Therefore, the probability in (41) approaches one as SNR tends to infinity.

Let us now analyze $Pr\left[\frac{\|\boldsymbol{h}'_l\|^2\|\boldsymbol{g}\|^2}{\|\boldsymbol{h}'_l\|^2\|\boldsymbol{g}\|^2 - \langle\boldsymbol{h}'_l, \boldsymbol{g}\rangle^2} < \text{SNR}^\epsilon\right]$. Whenever $\frac{\|\boldsymbol{h}'_l\|^2\|\boldsymbol{g}\|^2}{\|\boldsymbol{h}'_l\|^2\|\boldsymbol{g}\|^2 - \langle\boldsymbol{h}'_l, \boldsymbol{g}\rangle^2} < \text{SNR}^\epsilon$, it means that

$$\frac{\left|\langle\boldsymbol{h}'_l, \boldsymbol{g}\rangle\right|^2}{\|\boldsymbol{h}'_l\|^2\|\boldsymbol{g}\|^2} < 1 - \frac{1}{\text{SNR}^\epsilon}.$$

Let us consider the normalized vector $\underline{\boldsymbol{h}}'_l = \frac{\boldsymbol{h}'_l}{\|\boldsymbol{h}'_l\|}$ and its orthogonal normalized vector $(\underline{\boldsymbol{h}}'_l)^\perp = \frac{(\boldsymbol{h}'_l)^\perp}{(\|\boldsymbol{h}'_l\|)^\perp}$. The channel vector $\boldsymbol{g}$ can be decomposed into parallel and orthogonal components as follows.

$$\boldsymbol{g} = \langle\boldsymbol{g}, \underline{\boldsymbol{h}}'_l\rangle\underline{\boldsymbol{h}}'_l + \langle\boldsymbol{g}, (\underline{\boldsymbol{h}}'_l)^\perp\rangle(\underline{\boldsymbol{h}}'_l)^\perp.$$

We use $\boldsymbol{g}^\perp = \langle\boldsymbol{h}'_l, (\underline{\boldsymbol{h}}'_l)^\perp\rangle$ and $\boldsymbol{g}^\| = \langle\boldsymbol{g}, \underline{\boldsymbol{h}}'_l\rangle$ to denote the perpendicular and parallel components, respectively. Since

$$|C(z_{l,1}, z_{l,2})| = \begin{vmatrix} 2mP\sum_{j=1}^{L}|h'_{l,j}|^2 + \sigma^2 & 0 \\ 0 & 2mP\frac{(\sum_{j=1}^{L} h'_{l,j})^2}{\varrho^2}\sum_{j=1}^{L}|g'_j|^2 + \sigma^2 \end{vmatrix}$$
$$= \left(2mP\sum_{j=1}^{L}|h'_{l,j}|^2 + \sigma^2\right)\left(2mP\frac{(\sum_{j=1}^{L} h'_{l,j})^2}{\varrho^2}\sum_{j=1}^{L}|g'_j|^2 + \sigma^2\right) \quad (36)$$
$$= \left(2mP\sum_{j=1}^{L}|h'_{l,j}|^2 + \sigma^2\right)\left(2mP(\sum_{j=1}^{L} h'_{l,j})^2 + \sigma^2\right).$$

---

$$E[|C(z_{l,1}, z_{l,2}|x_{1,1}, x_{1,2})|]$$
$$= E\left[\begin{vmatrix} 2P((m-1)|h'_{l,1}|^2 + m\sum_{l=2}^{L}|h'_{l,j}|^2) + \sigma^2 & -\frac{(\sum_{j=1}^{L} h'_{l,j})^* x^*_{1,1}}{\varrho x^*_{1,2}}P\left(2(m-1)(g'_1)^* h'_{l,1} + 2m\sum_{j=2}^{L}(g'_j)^* h'_{l,j}\right) \\ \frac{(\sum_{j=1}^{L} h'_{l,j}) x_{1,1}}{\varrho x_{1,2}}P\left(2(m-1)(g'_1)(h'_{l,1})^* + 2m\sum_{j=2}^{L}(g'_j)(h'_{l,j})^*\right) & \frac{|x_{1,1}|^2}{|x_{1,2}|^2}P\frac{(\sum_{j=1}^{l} h'_{l,j})^2}{\varrho^2}\left(2(m-1)|g'_1|^2 + 2m\sum_{j=2}^{L}|g'_j|^2\right) + \sigma^2 \end{vmatrix}\right]$$
$$= \sigma^2\left(2P((m-1)|h'_{l,1}|^2 + m\sum_{j=2}^{L}|h'_{l,j}|^2) + 2P\frac{(\sum_{j=1}^{L} h'_{l,j})^2}{\varrho^2}\left((m-1)|g'_1|^2 + m\sum_{j=2}^{L}|g'_j|^2\right) + \sigma^2\right) + 4P^2\frac{(\sum_{j=1}^{L} h'_{l,j})^2}{\varrho^2}\left[\left((m-1)|g'_1|^2\right.\right.$$
$$\left.\left.+ m\sum_{j=2}^{L}|g'_j|^2\right)\left((m-1)|h'_{l,j}|^2 + 2m\sum_{j=2}^{L}|h'_{l,j}|^2\right) - \left|(m-1)(g'_1)^* h'_{l,1} + m\sum_{j=2}^{L}(g'_j)^* h'_{l,j}\right|^2\right]. \quad (37)$$

---

$\|\boldsymbol{g}\|^2 = \|\boldsymbol{g}^\perp\|^2 + \|\boldsymbol{g}^\parallel\|^2$, we can write

$$\frac{\left|\langle \boldsymbol{h}'_l, \boldsymbol{g}\rangle\right|^2}{\|\boldsymbol{h}'_l\|^2\|\boldsymbol{g}\|^2} = \frac{\left|\langle \frac{\boldsymbol{h}'_l}{\|\boldsymbol{h}'_l\|}, \boldsymbol{g}\rangle\right|^2}{\|\boldsymbol{g}\|^2}$$
$$= \frac{\left|\langle \underline{\boldsymbol{h}}'_l, \boldsymbol{g}\rangle\right|^2}{\|\boldsymbol{g}\|^2} \quad (42)$$
$$= \frac{\|\boldsymbol{g}^\parallel\|^2}{\|\boldsymbol{g}^\perp\|^2 + \|\boldsymbol{g}^\parallel\|^2}.$$

The magnitude of the elements of the channel vector $\boldsymbol{g}$ between the selected relays and the destination follows Rayleigh distribution. From [21], $\|\boldsymbol{g}^\parallel\|^2 \sim \Gamma(1,1)$ and $\|(\boldsymbol{g})^\perp\|^2 \sim \Gamma(L-1,1)$, where $\Gamma(p,\lambda)$ denotes the $\Gamma$ distribution with parameters $(p,\lambda)$. By applying this result and considering (42), we can obtain [22]

$$\frac{\left|\langle \boldsymbol{h}'_l, \boldsymbol{g}\rangle\right|^2}{\|\boldsymbol{h}'_l\|^2\|\boldsymbol{g}\|^2} \sim \beta(1, L-1),$$

where $\beta(p,\lambda)$ denotes the Beta distribution with parameters $(p,\lambda)$. Furthermore, it is known that the cumulative distribution function (CDF) of the Beta distribution is the regularized incomplete Beta function [22]. Consequently, we obtain,

$$P_r\left[\frac{\|\boldsymbol{h}'_l\|^2\|\boldsymbol{g}\|^2}{\|\boldsymbol{h}'_l\|^2\|\boldsymbol{g}\|^2 - \langle \boldsymbol{h}'_l, \boldsymbol{g}\rangle^2} < \text{SNR}^\epsilon, \ \forall l \in [L+1, K]\right]$$
$$= \prod_{l=L+1}^{K} P_r\left[\frac{\|\boldsymbol{h}'_l\|^2\|\boldsymbol{g}\|^2}{\|\boldsymbol{h}'_l\|^2\|\boldsymbol{g}\|^2 - \langle \boldsymbol{h}'_l, \boldsymbol{g}\rangle^2} < \text{SNR}^\epsilon\right]$$
$$= \left[P_r\left(\frac{\|\boldsymbol{h}'_{L+1}\|^2\|\boldsymbol{g}\|^2}{\|\boldsymbol{h}'_{L+1}\|^2\|\boldsymbol{g}\|^2 - \langle \boldsymbol{h}'_{L+1}, \boldsymbol{g}\rangle^2} < \text{SNR}^\epsilon\right)\right]^{K-L}$$
$$= \left(P_r\left[\frac{\left|\langle \boldsymbol{h}'_{L+1}, \boldsymbol{g}\rangle\right|^2}{\|\boldsymbol{h}'_{L+1}\|^2\|\boldsymbol{g}\|^2} < 1 - \frac{1}{\text{SNR}^\epsilon}\right]\right)^{K-L}$$
$$= \left(\frac{\beta\left(1 - \frac{1}{\text{SNR}^\epsilon}; 1, L-1\right)}{\beta(1, L-1)}\right)^{K-L}$$
$$= \left(I_{1-\frac{1}{\text{SNR}^\epsilon}}(1, L-1)\right)^{K-L}$$
$$= \left(1 - \left(\frac{1}{\text{SNR}^\epsilon}\right)^{L-1}\right)^{K-L}$$
$$\underset{\text{SNR}\to\infty}{\to} 1, \quad (43)$$

where $\beta\left(1 - \frac{1}{\text{SNR}^\epsilon}; 1, L-1\right)$ and $I_{1-\frac{1}{\text{SNR}^\epsilon}}(1, L-1)$ denotes the incomplete Beta function and the regularized incomplete Beta function, respectively. The probabilities in (41) and (43) approach one as SNR tends to infinity. There product is equal to $P_{out1}(\text{SNR}) \underset{\text{SNR}\to\infty}{\to} 1$. This proves (22).

## APPENDIX E

$$P_{out}(\gamma, \text{SNR})$$
$$\leqslant P_r \left[ R_{\text{Bob}} - \frac{1}{2}\log_2(\text{SNR}^\epsilon) < \gamma \Big| \max_{l\in[1,K]} R_l < \frac{1}{2}\log_2(\text{SNR}^\epsilon) \right]$$
$$= P_r \left[ \frac{1}{2}\log_2\left(1 + \frac{\text{SNR}\sum_{l=1}^L |g_l|^2}{\left(1 + \sum_{l=1}^L \frac{|g_l|^2}{|h'_{l,l}|^2}\right)}\right) - \frac{1}{2} - \frac{1}{2}\log_2(\text{SNR}^\epsilon) < \gamma \right.$$
$$\left. \Big| \max_{l\in[1,K]} R_l < \frac{1}{2}\log_2(\text{SNR}^\epsilon) \right]$$
$$\leqslant P_r \left[ \frac{1}{2}\log_2\left(\frac{\text{SNR}^{1-\epsilon}\sum_{l=1}^L |g_l|^2}{1 + \sum_{l=1}^L \frac{|g_l|^2}{|h'_{l,l}|^2}}\right) < \gamma + \frac{1}{2} \right.$$
$$\left. \Big| \max_{l\in[1,K]} R_l < \frac{1}{2}\log_2(\text{SNR}^\epsilon) \right]. \quad (44)$$

whenever $R_{\text{Bob}} - \frac{1}{2}\log_2(\text{SNR}^\epsilon) < \gamma$ and $\max_{l\in[1,K]} R_l < \frac{1}{2}\log_2(\text{SNR}^\epsilon)$ yield

$$\begin{cases} \frac{\text{SNR}^{1-\epsilon}\sum_{l=1}^L |g_l|^2}{1+\sum_{l=1}^L \frac{|g_l|^2}{|h'_{l,l}|^2}} < 2^{2\gamma+1} \\ |g_{lm}|^2 < (\text{SNR}^\epsilon - 1)\sum_{\substack{j=1\\j\neq lm}}^L |g_j|^2 \end{cases}. \quad (45)$$

This is equivalent to the following inequalities.

$$\begin{cases} \text{SNR}^{1-\epsilon}\sum_{l=1}^L |g_l|^2 - 2^{2\gamma+1}\sum_{l=1}^L \frac{|g_l|^2}{|h'_{l,l}|^2} < 2^{2\gamma+1} \\ |g_{lm}|^2 < (\text{SNR}^\epsilon - 1)\sum_{\substack{j=1\\j\neq lm}}^L |g_j|^2, \end{cases} \quad (46)$$

which is equivalent to

$$\begin{cases} \text{SNR}^{1-\epsilon}|g_{lm}|^2 - \frac{2^{2\gamma+1}|g_{lm}|^2}{|h'_{lm,lm}|^2} + \sum_{\substack{l=1\\l\neq lm}}^L |g_l|^2\left(\text{SNR}^{1-\epsilon} - \frac{2^{2\gamma+1}}{|h'_{l,l}|^2}\right) < 2^{2\gamma+1} \\ |g_{lm}|^2 < (\text{SNR}^\epsilon - 1)\sum_{\substack{j=1\\j\neq lm}}^L |g_j|^2. \end{cases} \quad (47)$$

To obtain an upper bound on the outage probability at high SNR, we provide a lower bound on the first inequality in (47). This can be obtained by subtracting the first term, in the first inequality, and replacing $|g_{lm}|$ in the second term, in the first inequality, by its upper bound provided by the second inequality in (47). That is,

$$-\frac{2^{2\gamma+1}\sum_{\substack{l=1\\l\neq lm}}^L |g_l|^2}{|h'_{lm,lm}|^2} + \sum_{\substack{l=1\\l\neq lm}}^L |g_l|^2\left(\text{SNR}^{1-\epsilon} - \frac{2^{2\gamma+1}}{|h'_{l,l}|^2}\right) < 2^{2\gamma+1}$$

which can be rearranged as

$$\sum_{\substack{l=1\\l\neq lm}}^L |g_l|^2 \left(\text{SNR}^{1-\epsilon} - 2^{2\gamma+1}\left(\frac{1}{|h'_{l,l}|^2} + \frac{1}{|h'_{lm,lm}|^2}\right)\right) < 2^{2\gamma+1}. \quad (48)$$

At high SNR, the outage probability can thus be outer bounded as given in (49) on the next page.

The probability $P_{out2}(\text{SNR}^\epsilon) = P_r \left[ \max_{l\in[1,L]} \frac{1}{|h'_{l,l}|^2} < \text{SNR}^\epsilon \right]$ approaches one as SNR tends to infinity. Indeed, the magnitude of the channel vector $h'_l$ resulting from ZFBF follows Rayleigh distribution [18]. Therefore $P_{out2}(\text{SNR}^\epsilon)$ can be written as follows.

$$P_r \left[ \max_{l\in[1,L]} \frac{1}{|h'_{l,l}|^2} < \text{SNR}^\epsilon \right] = P_r \left[ \frac{1}{|h'_{l,l}|^2} < \text{SNR}^\epsilon, \forall l \in [1,L] \right]$$
$$= \left( P_r \left[ \frac{1}{|h'_1|^2} < \text{SNR}^\epsilon \right] \right)^L$$
$$= \left( 1 - P_r \left[ |h'_1|^2 > \frac{1}{\text{SNR}^\epsilon} \right] \right)^L$$
$$= \left( 1 - e^{\frac{1}{2\text{SNR}^\epsilon}} \right)^L$$
$$\underset{\text{SNR}\to\infty}{\to} 0. \quad (50)$$

This yields at high SNR an upper bound on the outage probability as

$$P_{out}(\gamma, \text{SNR}) \leqslant P_r \left[ \sum_{\substack{l=1\\l\neq lm}}^L |g_l|^2 \left(\text{SNR}^{1-\epsilon} - \frac{2^{2\gamma+1}}{|h'_{l,l}|^2} + \frac{2^{2\gamma+1}}{|h'_{lm}|^2}\right) < 2^{2\gamma+1} \right.$$
$$\left. \Big| \max_{l\in[1,L]} \frac{1}{|h'_{l,l}|^2} < \text{SNR}^\epsilon \right]$$
$$\leqslant P_r \left[ \sum_{\substack{l=1\\l\neq lm}}^L |g_l|^2 \left(\text{SNR}^{1-\epsilon} - 2^{2\gamma+1}(2\text{SNR}^\epsilon)\right) < 2^{2\gamma+1} \right]$$
$$= P_r \left[ \sum_{\substack{l=1\\l\neq lm}}^L |g_l|^2 < \frac{2^{2\gamma+1}}{\left(\text{SNR}^{1-\epsilon} - 2^{2\gamma+2}\text{SNR}^\epsilon\right)} \right]. \quad (51)$$

$$\begin{aligned}
&P_{out}(\text{SNR}, \gamma) \\
&\leqslant P_r\left[\sum_{\substack{l=1 \\ l\neq lm}}^{L}|g_l|^2\left(\text{SNR}^{1-\epsilon} - 2^{2\gamma+1}\left(\frac{1}{|h'_{l,l}|^2}+\frac{1}{|h'_{lm}|^2}\right)\right) < 2^{2\gamma+1}\right] \\
&= \left(1 - P_r\left[\max_{l\in[1,L]}\frac{1}{|h'_{l,l}|^2} < \text{SNR}^{\epsilon}\right]\right) P_r\left[\sum_{\substack{l=1 \\ l\neq lm}}^{L}|g_l|^2\left(\text{SNR}^{1-\epsilon} - 2^{2\gamma+1}\left(\frac{1}{|h'_{l,l}|^2}+\frac{1}{|h'_{lm}|^2}\right)\right) < 2^{2\gamma+1}\Big|\max_{l\in[1,L]}\frac{1}{|h'_{l,l}|^2} \geqslant \text{SNR}^{\epsilon}\right] \\
&\quad + P_r\left[\max_{l\in[1,L]}\frac{1}{|h'_{l,l}|^2} < \text{SNR}^{\epsilon}\right] P_r\left[\sum_{\substack{l=1 \\ l\neq lm}}^{L}|g_l|^2\left(\text{SNR}^{1-\epsilon} - 2^{2\gamma+1}\left(\frac{1}{|h'_{l,l}|^2}+\frac{1}{|h'_{lm}|^2}\right)\right) < 2^{2\gamma+1}\Big|\max_{l\in[1,L]}\frac{1}{|h'_{l,l}|^2} < \text{SNR}^{\epsilon}\right]
\end{aligned} \quad (49)$$